\documentclass[aps,prl,twocolumn,showpacs,superscriptaddress,footinbib,preprintnumbers]{revtex4-1}

\usepackage{style}
\usepackage{colortbl}
\usepackage{multirow}
\usepackage{graphicx}
\usepackage{amsmath}
\usepackage{amssymb}
\usepackage{amsthm}
\usepackage{amsfonts}
\usepackage{braket}
\usepackage{slashed}
\usepackage[T1]{fontenc}
\usepackage{mathtools}

\definecolor{newgray}{gray}{0.95}

\newcommand{\lr}[1]{\left( #1\right)}

\begin{document}

\preprint{KEK-TH-2588}
\preprint{KEK-Cosmo-0336} 
\preprint{KEK-QUP-2023-0038}
\preprint{IPMU23-0051}
\preprint{IFIC/23-54}
\preprint{ULB-TH/23-18} 

\title{Non-Canonical Nucleon Decays as Window into Light New Physics}

\author{K\r{a}re Fridell}
\email{kare.fridell@kek.jp}
\affiliation{Theory Center, Institute of Particle and Nuclear Studies, High Energy Accelerator Research Organization (KEK), Tsukuba 305-0801, Japan}
\affiliation{Department of Physics, Florida State University, Tallahassee, FL 32306, USA}

\author{Chandan Hati} 
\email{chandan@ific.uv.es}
\affiliation{Instituto de Física Corpuscular (IFIC), Universitat de València-CSIC,
C/ Catedratico Jose Beltran, 2, E-46980 Valencia, Spain}
\affiliation{Service de Physique Th\'eorique, Universit\'e Libre de Bruxelles,
Boulevard du Triomphe, CP225, 1050 Brussels, Belgium}

\author{Volodymyr Takhistov}
\email{vtakhist@post.kek.jp}
\affiliation{International Center for Quantum-field Measurement Systems for Studies of the Universe
and Particles (QUP), KEK, 1-1 Oho, Tsukuba, Ibaraki 305-0801, Japan
}
\affiliation{Theory Center, Institute of Particle and Nuclear Studies, High Energy Accelerator Research Organization (KEK), Tsukuba 305-0801, Japan}
\affiliation{Graduate University for Advanced Studies (SOKENDAI),
1-1 Oho, Tsukuba, Ibaraki 305-0801, Japan
}
\affiliation{Kavli Institute for the Physics and Mathematics of the Universe (WPI), Chiba 277-8583, Japan}

\begin{abstract} 
\noindent Nucleon decays are generic predictions of motivated theories, including those based on the unification of forces and supersymmetry. We demonstrate that non-canonical nucleon decays offer a unique opportunity to 
broadly probe light new particles beyond the Standard Model with masses below $\sim$few~GeV over decades in mass range, including
axion-like particles, dark photons, sterile neutrinos, and scalar dark matter. 
Conventional searches can misinterpret and even completely miss such new physics.
We propose a general strategy based on momenta of visible decay final states to probe these processes,
offering a rich physics program for existing and upcoming experiments such as Super-Kamiokande,
Hyper-Kamiokande, DUNE, and JUNO.
\end{abstract}

\maketitle

\paragraph{\textbf{Introduction.} ---} 
Baryon number $B$ appears to be an accidentally conserved symmetry of the Standard Model (SM) that ensures the stability of protons. However, many considerations strongly motivate and highlight the fundamental necessity of searching for observable $B$-violating (BNV) $\Delta B \neq 0$ processes that would constitute a clear sign of new physics beyond the SM.
While strongly suppressed at low temperatures, $B$ is already violated by three units in the SM through non-perturbative instanton effects~\cite{tHooft:1976rip}. Global symmetries in general, including $B$ or lepton number $L$, are expected to be broken by quantum gravity effects~\cite{Banks:2010zn,Harlow:2018jwu}. Violation of $B$ is one of the key conditions necessary to successfully generate observed asymmetry between baryons and anti-baryons in the early Universe~\cite{Sakharov:1967dj}. Nucleon decays, corresponding to BNV processes, naturally appear within the context of fundamental Grand Unified Theories (GUTs)~\cite{Georgi:1974sy,Fritzsch:1974nn} and theories based on supersymmetry (see Refs.~\cite{Langacker:1980js,Nath:2006ut} for a review).  

Significant efforts have been devoted to searching for fundamental BNV nucleon decays, with over 60 processes already analyzed over the span of several decades~\cite{Workman:2022ynf}. Diverse strategies for analyzing such processes have been discussed~\cite{Heeck:2019kgr,Dev:2022jbf}. Most sensitive limits have been obtained by the Super-Kamiokande large water Cherenkov experiment~\cite{Super-Kamiokande:2002weg} (see Ref.~\cite{Takhistov:2016eqm} for a review), pushing the nucleon lifetime above $\sim 10^{34}$~years for proton decays $p\rightarrow e^+\pi^0$, $p\rightarrow \mu^+\pi^0$~\cite{Super-Kamiokande:2020wjk} and ruling out or constraining broad classes of theories, especially minimal GUT models. Nucleon decays constitute prime physics targets for upcoming experiments including Hyper-Kamiokande~\cite{Hyper-Kamiokande:2018ofw}, DUNE~\cite{DUNE:2020lwj} and JUNO~\cite{JUNO:2015sjr}.

In conventional nucleon decay searches it has been assumed that the observable processes only involve SM fields as the external final states. However, in a variety of theories nucleon decay can be induced by external interactions~\cite{Davoudiasl:2010am,Berger:2023ccd}, resulting in unusual kinematics of outgoing particles. In Ref.~\cite{Helo:2018bgb} it was suggested that $p \rightarrow (e^+ + {\rm missing  ~energy})$, combined with non-observation of $p \rightarrow e^+\pi^0$, could be related to the appearance of sterile neutrino. Invisible nucleon decays $n \rightarrow$ (invisible) have been discussed in the context of new dark sector fermions related to the neutron lifetime puzzle~\cite{Barducci:2018rlx,Fornal:2018eol,McKeen:2018xwc} and unparticles that do not behave like particles~\cite{He:2008rv}.

In this work, we demonstrate that non-canonical nucleon decays (NCNDK) constitute an underexplored frontier covering a broad landscape of motivated light new physics targets beyond the SM spanning decades of orders of magnitude in mass range below $\sim$few GeV, including scalars (e.g.\ axion-like particles (ALPs), majorons), neutral fermions (e.g.\ sterile neutrinos, light neutral composite fields), and gauge bosons (e.g.\ dark photons). We show that ongoing and upcoming experiments looking for the conventional nucleon decay modes are sensitive with distinguishable signatures to a plethora of novel scenarios where the nucleons can decay into different light states beyond the SM. We propose a general strategy to exploit these opportunities. 

\begin{table*}[t]
    \centering
    \setlength{\extrarowheight}{2pt}
    \begin{tabular}{m{14mm}  m{42mm}   
    c c l l}
    \hline
    \specialrule{1pt}{1pt}{0pt}
   $\mathcal{O}$  & Operator & $(\Delta B, \Delta L)$ & Dim & Decay modes & New Field(s) \\ 
   [1mm]
   \hline
   \multirow{1}*{$\mathcal{O}_{d^2 u N}$} & \multirow{1}*{$\epsilon^{abc} \left(\bar{d}^c_{a} N\right) \left(\bar{d}^c_{b} u_{c}\right)$} & \multirow{1}*{$(1,1)$} & \multirow{1}*{$6$} & $p (n)\to \pi^{+(0)} \bar N$ & sterile neutrino  \\[1mm]
    \hline
   \multirow{2}*{\raisebox{0mm}{$\mathcal{O}_{D d^2 u \bar N}$}} & \multirow{2}*{$\epsilon^{abc} \left(\bar{N} \gamma_{\mu }d_{a}\right) \left(\bar{d}_{b}^c D^{\mu } u_c\right)$} & \multirow{2}*{$(1,-1)$ }& \multirow{2}*{$7$} & $n\to  N\gamma$ & \multirow{2}*{sterile neutrino} \\
   & & & & $p (n)\to \pi^{+(0)} N\gamma$ &  \\[1mm]
    \hline
   \multirow{2}*{\raisebox{0mm}{$\mathcal{O}_{d u^2 e\phi}$}} & \multirow{2}*{$\epsilon^{abc}\left(\bar{d}^c_a u_b\right)\left(\bar{e}^c u_c\right)\phi^\dagger$} & \multirow{2}*{$(1,1)$} & \multirow{2}*{$7$} & $p\to e^+\phi$ & \multirow{2}*{dark scalar, majoron}\\
   & & & & $p (n)\to e^+\pi^{0(-)}\phi$ &  \\[1mm] 
    \hline
   \multirow{2}*{\raisebox{0mm}{$\mathcal{O}_{d^2 Q \bar L X}$}} &\multirow{2}*{$\epsilon^{abc}\left( \bar{Q}^c_a{}^{i}\gamma_\mu d_{b}\right)\big(\bar L_i d_{c}\big)X^\mu$} & \multirow{2}*{$(1,-1)$} & \multirow{2}*{$7$} & $n\to \nu X\; / \; e^-\pi^+ X$ &  \multirow{2}*{dark photon} \\
   & & & & $p (n) \to \nu\pi^{+(0)} X$ &    \\[1mm]
    \hline
   \raisebox{0mm}{$\mathcal{O}_{d Q^2 \bar L \bar H\phi}$} & $\epsilon^{abc}\left(\bar{Q}^c_a{}^iQ_{b}^j\right)\left(\bar{L}_{i}d_c\right)H_j^\dagger\phi^\dagger$ & $(1,-1)$ & $8$ & $n\to \nu\phi\; /\; e^-\pi^{+}\phi$ & dark scalar, majoron\\[1mm] 
    \hline
   \raisebox{0mm}{$\mathcal{O}_{D d^2 Q \bar L a}$} & ${\epsilon^{abc}}({\partial_{\mu}}a)\left(\bar{Q}^c_a{}^{i}\gamma^\mu d_{b}\right)(\bar{L}_i{d_c}) $ & $(1,-1)$ & $8$ & $n\to \nu  a\; /\; e^-\pi^{+} a$ & axion-like particles \\[1mm]
    \hline
   \multirow{2}*{\raisebox{0mm}{$\mathcal{O}_{D d^2 u \bar N a}$}} & \multirow{2}*{${\epsilon^{abc}}({\partial_{\mu}}a)\left(\bar{N} \gamma^{\mu }d_{a}\right) \left(\bar{d}^c_{b} u_c\right)$} & \multirow{2}*{$(1,-1)$} & \multirow{2}*{$8$} & $n\to Na$ & \raisebox{-1mm}{axion-like particle with}\\
   & & & & $p (n)\to \pi^{+(0)} Na$ & \multirow{1}*{sterile neutrino} \\[1mm]
    \hline
   \multirow{2}*{$\mathcal{O}_{d u Q e \bar L \bar N}$} & \multirow{2}*{$\epsilon^{abc}\left(\bar{e}^c u_a\right)\left(\bar{Q}^c_b{}^i\gamma_\mu d_c\right)\big(\bar{L}_i\gamma^\mu N^c\big)$} & \multirow{2}*{$(1,-1)$} & \multirow{2}*{$9$} & $p\to e^+\nu N$ & \multirow{2}*{sterile neutrino} \\
   & & & & $n \to e^+e^- N$   \\[1mm]
    \hline
   $\mathcal{O}_{d u^2 e N^2}$ & $\epsilon^{abc}\left(\bar{d}^c_a u_b\right)\left(\bar{e}^c u_c\right)(\bar{N}^c N)$ & $(1,3)$ & $9$ & $p\to e^+\bar N\bar N$ & sterile neutrino \\[1mm]
      \specialrule{1pt}{0pt}{2pt}
      \hline
    \end{tabular}
    \caption{Characteristic list of SM invariant $B$-violating operators mediating NCNDKs in four-component spinor notation involving the SM as well as additional new light degrees of freedom. Here, $\phi$ denotes new light scalars, $N$ fermions, $X$ vector fields, and $a$ ALPs, all being singlets under the SM gauge group. Processes are shown only up to 3-body decays.
    }
    \label{tab:op_list}
\end{table*}
 
\paragraph{\textbf{Novel nucleon decays with light states.} --- } 

Nucleon decay processes mediated by $\Delta B \neq 0$ interactions can be explored in generality using
SM effective field theory (SMEFT)~\cite{Weinberg:1979sa,Wilczek:1979hc,Abbott:1980zj,Beneito:2023xbk}.
From the low energy SM perspective, $\Delta B = 1$ operators start to manifest at the lowest order at dimension six ($d=6$). 
Particularly well studied are $|\Delta (B-L)| = 0$ conserving processes, such as $p \rightarrow e^+ \pi^0$ that often is the dominant channel of non-supersymmetric theories and sensitive to GUT scale physics around $\sim10^{16}$~GeV.
At $d=7$, for example, there are $|\Delta (B-L)| = 2$ nucleon decays~\cite{Weinberg:1980bf,Weldon:1980gi} such as $n \rightarrow e^- \pi^+$ discussed in context of
the Pati-Salam model and $SO(10)$ GUTs~\cite{Pati:1983zp,Babu:2012vb,Hati:2018cqp}. 
Additional channels from higher dimensions, such as $p\rightarrow e^+e^+\mu^-$, can be observable if the scale of new physics mediating the processes is significantly below GUT scales~\cite{ODonnell:1993kdg,Hambye:2017qix,Fonseca:2018ehk}. A systematic overview of nucleon decay modes with an emphasis on new inclusive searches can be found in Ref.~\cite{Heeck:2019kgr}.

To elucidate light new physics scenarios that can lead to NCNDKs, we will consider effective interactions involving scalars, pseudo-scalars, neutral fermions, and gauge bosons. These possibilities can be taken to correspond to well-motivated and actively searched-for scenarios of light new physics beyond the SM, such as dark scalars that may contribute to dark matter or majorons that are Goldstone bosons associated with the lepton doublet $L$~\cite{Chikashige:1980ui,Gelmini:1980re}, ALPs that are general (pseudo) Goldstone bosons of a global $U(1)$ symmetry breaking (e.g.\ Ref.~\cite{Adams:2022pbo}), sterile neutrinos that mix with active SM neutrinos~(e.g.\ Ref.~\cite{Boyarsky:2018tvu}), light neutral composite fields~\cite{Arkani-Hamed:1998wff,Agashe:2015izu,Chacko:2020zze,Chakraborty:2021fkp}, or dark photon gauge bosons~(e.g.\ Ref.~\cite{Fabbrichesi:2020wbt}). 

\begin{figure*}[t!]
    \centering  \includegraphics[width=0.45\textwidth]
{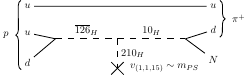}
    \hspace{3em}
\raisebox{2mm}{\includegraphics[width=0.43\textwidth]{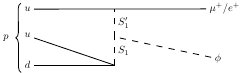}}
    \caption{[Left] Proton decay $p \rightarrow \pi^+N$ in the GUT model described in the text, with representations shown for $SO(10)$ GUT, and the vacuum expectation value for the Pati-Salam gauge group $SU(2)_L\times SU(2)_R\times SU(4)_C$. [Right] Proton decay $p \rightarrow e^+(\mu^+)\phi$ mediated by the $S_1$ and $S_1'$ leptoquarks. See text for details.}
    \label{fig:diagrams}
\end{figure*}

In particular, we illustrate such characteristic new physics interactions focusing on the operators up to dimension 9 listed in Tab.~\ref{tab:op_list}. Furthermore, higher dimensional operators can also be constructed similarly~\footnote{At higher dimensions nucleon decays can occur at loop level as well, allowing for the scenarios to be realized with new physics.}. As depicted in the last columns of Tab.~\ref{tab:op_list}, different operators will lead to one or multiple missing energy contributions in the nucleon decay final states. Intriguingly, we note that novel nucleon decay modes $n \rightarrow N a$ (i.e. an ALP with a sterile neutrino) as well as $n \rightarrow \nu X$ (i.e. a neutrino with a dark photon) constitute the most minimal invisible nucleon decay channels, since considering only SM final states gives $n \rightarrow \nu\nu\nu$. Invisible nucleon decays have been searched in various experiments~\cite{SNO:2018ydj,KamLAND:2005pen} and could become significant in models based on extra dimensions (see e.g.\ Refs.~\cite{Dvali:1999hn,Mohapatra:2002ug}) or partially unified Pati–Salam type theories~\cite{Pati:1973rp}.

We stress that the characteristic list of our NCNDK modes in Tab.~\ref{tab:op_list} is not exhaustive. Dinucleon (see e.g.\ Refs.~\cite{Super-Kamiokande:2015pys,Super-Kamiokande:2018apg} for dinucleon decays with SM final states) and trinucleon decays (see e.g.\ Ref.~\cite{Majorana:2018pdo} for trinucleon decays with SM final states), and new radiative modes analogous to $p \rightarrow e^+\gamma$ with SM final states~\cite{Silverman:1980ha} are also possible. More so, as exemplified by $n \rightarrow Na$, a broad variety of additional NCNDKs involving a combination of novel light final state particles that can have distinct theoretical motivations are feasible. We leave a comprehensive systematic analysis of possible NCNDK channels for future work.

The two-body NCNDK widths corresponding to the operators in Tab.~\ref{tab:op_list} are given by
\begin{equation}
\begin{aligned}
    \Gamma_{\psi\to i j} = \frac{1}{16\pi}\frac{\lambda^{1/2}(m_\psi,m_{i},m_{j})}{m_\psi^3}|\sum_{I}C_I \mathcal{M}_I^{\psi\to i j}|^2 \, ,
\end{aligned}
\end{equation}
with $\lambda$ being the K\"all\'en function $\lambda (x,y,z) = x^2 + y^2 + z^2 - 2xy - 2yz - 2zx$. Here the Wilson coefficient $C_I$ corresponding to the operator $O_I$ is obtained by integrating out all the heavy degrees of freedom mediating NCNDK in a UV completion at the heavy new physics scale. The matrix element $\mathcal{M}_I$ is evaluated taking the values provided from Lattice QCD simulations, e.g.\ Ref.~\cite{Aoki:2017puj} at a typical hadronic scale $\mu_0=2$ GeV and then taking into account the running of the operators between $\mu_0$ and the heavy new physics scale. 

%
For the 3-body nucleon decay, the differential rate for $\psi\to i j k$ is given by
\begin{equation}
    \frac{d\Gamma_{\psi\to i j k}}{d|\vec p_i|}= \frac{1}{(2\pi)^3}\frac{1}{32m_\psi^3} \frac{2m_\psi |\vec p_i|}{\sqrt{m_i^2+|\vec p_i|^2}} \int_{t^-}^{t^+} dt |\mathcal{M}_{\psi\to i j k}|^2~,
\end{equation}
where $s\equiv (p_{\psi}-p_i)^2$ and $t\equiv (p_\psi-p_j)^2$. A concise summary of the details of the kinematics and the matrix element calculation, including the form factor formalism and renormalization group running, are provided in the Appendix. 

\paragraph{\textbf{Ultraviolet model completion.} ---}  We now illustrate how two characteristic distinct processes from Tab.~\ref{tab:op_list}, based on $\mathcal{O}_{d^2 u N}$ leading to $p\to N\pi^+$ and $\mathcal{O}_{d u^2 LH\phi}$ leading to $p\to e^+\phi$, can be realized within concrete models. For $p\to N\pi^+$, the sterile state $N$ must have a mass below the proton mass. Further, given our BNV effective interaction, a natural context to consider it in are the left-right symmetric models~\cite{Mohapatra:1974gc,Senjanovic:1975rk,Mohapatra:1979ia,Mohapatra:1980yp} with the gauge group $G_{LR}\in SU(3)_C \times SU(2)_L\times SU(2)_R\times U(1)_{B-L}$. If the masses of right-handed neutrinos $N$ are significantly lighter than the right-handed gauge bosons, and the right-handed gauge bosons lie well above the electroweak scale, then the feebleness of the right-handed interactions make $N$ effectively ``sterile'' states. 

To mediate the $p\to N\pi^+$ mode, we need mediators with diquark and leptoquark couplings, which are well known to be present in theories unifying leptons and quarks, like the Pati-Salam model~\cite{Pati:1974yy} (based on the gauge group $G_{PS}\in SU(2)_L\times SU(2)_R \times SU(4)_C $, which can be embedded in $SO(10)$). However, in the conventional minimal non-supersymmetric $SO(10)$ GUT models, the right-handed neutrino mass $m_N$ is often intimately linked with $SU(2)_R$ breaking occurring at relatively high scales $\mathcal{O}(10^{12-14})$~GeV as necessary to realize the conventional high-scale seesaw. Therefore $p\to N\pi^+$ is a challenge to realize in this context. 

The related conventional nucleon decay with SM final states $p\to\pi^+ \nu$ can be mediated via the leptoquark and diquark type mediators (in the $\overline{126}_H$ multiplet) with masses close to the GUT scale. However, the resultant dimension-6 operators are significantly suppressed because of the GUT scale mediators. Sizeable proton decay rates can be obtained in the supersymmetric version of this model~\cite{Goh:2003nv,Babu:2018tfi}. 

Implementation of $SO(10)$ breaking through the Pati-Salam route with an explicit $D$-parity breaking~\cite{Chang:1983fu,Chang:1984uy,Deppisch:2014qpa,Deppisch:2014zta} can lead to light enough $m_N$ realizing $p\to N\pi^+$ at tree level without supersymmetry~\footnote{Some of the other possibilities could include implementation of special symmetries in supersymmetric version of Pati-Salam model~\cite{King:1997wf} or use of extended seesaw mechanisms~\cite{Dolan:2020doe}.}. 
The most salient features regarding this scenario are the following--- (i) The $SO(10)$ breaking scale $m_U$, $D$-parity breaking scale $m_P$, the Pati-Salam symmetry breaking scale $m_{PS}$, $SU(2)_R$ (in $G_{LR}$) symmetry breaking scale $m_{R}$, and $B-L$ symmetry breaking scale $m_{B-L}$ can be decoupled: $m_U \gg m_P> m_{PS} \gg m_R >m_{B-L}\gg m_{SM}$. (ii) For type-II seesaw neutrino mass dominance the active neutrino mass is given by $m_\nu\simeq \beta v^2 m_R/ M\langle\sigma \rangle$, where $\beta$ is $\mathcal{O}(1)$ constant, $v\sim 246$ GeV is the electroweak vacuum expectation value, and $M$ and $\sigma$ are of the order of $m_P$~\cite{Sarkar:2004hc,Gong:2007yv}. (iii) For $m_P\gg m_R\sim m_{W_R}$, this allows for a decoupling of masses $m_N \ll m_R$, in contrast to the case of type-I seesaw scenario. 

An example diagram for the $p\to N\pi^+$ decay within this model is shown in Fig.~\ref{fig:diagrams}. The resulting proton lifetime is given by
\begin{equation}
\begin{aligned}
    \frac{\tau_{p\to N\pi^+}}{8.1\!\times\! 10^{34} \text{y}} =&\frac{(\frac{m_N^2}{\text{GeV}^2}+0.86)^{-1}(\frac{m_{\overline{126}_H}}{2\times10^{8}\text{ GeV}})^4(\frac{m_{10_H}}{2\times10^{8}\text{ GeV}})^4}{\frac{\lambda^{1/2}(m_p^2,m_N^2,m_{\pi^+}^2)}{\text{GeV}^2}\lambda_{dN}^2\lambda_{ud}^2 (\frac{v_{(1,1,15)}}{2\times 10^8\text{ GeV}})^4}\, .
\end{aligned}
\end{equation}
where $\lambda_{dN}$ and $\lambda_{ud}$ are the Yukawa couplings in Fig.~\ref{fig:diagrams} (left). Since the diquark and leptoquark mediator masses are generated when $SU(4)_C$ is broken to $U(1)_{B-L}\times SU(3)_C$ at the scale $m_{PS}$ via the Pati-Salam multiplet $(1,1,15) \subset 210_H$, their masses are of order $m_{PS}$ instead of $m_U$ as in the conventional case. As a benchmark, assuming $m_N \sim 400$~MeV and $m_{PS}\sim 2 \times 10^8$~GeV, we find $\tau_{p\to N\pi^+} \sim  1.1 \times 10^{35}$ years, which could be within the reach of Hyper-K~\cite{Hyper-Kamiokande:2018ofw}. 

The effective interaction $\mathcal{O}_{d u^2 e\phi}$ leading to $p\to e^+\phi$ can be realized at tree level, as shown in Fig.~\ref{fig:diagrams}, within minimal ultraviolet extensions of the SM. One such scenario comprises two copies of the scalar leptoquark $S_1,\, S_1': (\bar{3},1,1/3)$. 
The relevant interactions are
\begin{eqnarray}
\mathcal{L}\supset && y_{S_1}\epsilon u^{c}_R d^{c}_R S_{1} -y_{S_1'}d^{c}_R e_{R}^{c} S_1' - \mu_1 \phi S_1' S^{\dagger}_1~. 
\end{eqnarray}
Here, $\epsilon$ corresponds to the $SU(3)_C$ antisymmetric tensor. 
Rapid proton decays at dimension-6, mediated via the coexistence of leptoquark and diquark couplings of $S_1$, in all generality, are absent in case $S_1$ only has diquark couplings, while $S_1'$ only has leptoquark couplings. When the above trilinear coupling $\mu_1$ is instead considered, the combination of the $S_1$ and $S_1'$ leptoquarks can lead to experimentally interesting proton decay lifetime, given by
\begin{equation}
\begin{aligned}
     \frac{\tau_{p\to \mu^+\phi}}{4.0\!\times\! 10^{34}\text{y}} =&\frac{\text{GeV}^2}{\lambda^{1/2}(m_p^2,m_\mu^2,m_{\phi}^2)} \frac{(\frac{m_{S_1}}{10^{15}\text{ GeV}})^4 (\frac{m_{S_1'}}{10^5\text{ GeV}})^4}{y_{S_1}^2y_{S_1'}^2(\frac{\mu_1}{10^{15} \text{ GeV}})^2}\, ,
\end{aligned}
\end{equation}
As a benchmark, taking $m_\phi\sim 700$~MeV, $m_{S_1'}\sim 100$~TeV, $\mu_1\sim m_{S_1}\sim 10^{15}$~GeV, and couplings $y_{S_1}\sim y_{S_1'}\sim 0.4$, we find $\tau_{p\to \mu^+\phi} \sim 1.1\times 10^{35}$ years.

\paragraph{\textbf{Observational strategy and visible momenta.} --- } Traditionally, nucleon decay searches have focused on processes where the decay products are a combination of the SM photon, charged leptons, light mesons, or missing energy associated with active neutrinos, depending on the process.
For these modes, the expected number of events after background subtraction shows a peaked distribution in the momentum of the observabled particle(s), as highlighted e.g.\ by spectral searches at Super-Kamiokande~\cite{Super-Kamiokande:2013rwg,Super-Kamiokande:2014pqx,Super-Kamiokande:2015pys}.

For novel NCNDKs with light but not massless new particles in the final state, the momenta distributions of experimentally visible SM final state constitutes can significantly differ w.r.t.\ the conventional nucleon decays, with distributions peaking at lower momenta compared to analogous modes with nearly massless neutrinos. Here, we do not distinguish neutrinos and anti-neutrinos.

\begin{figure}[t!]
    \centering
    \includegraphics[width=0.47\textwidth]{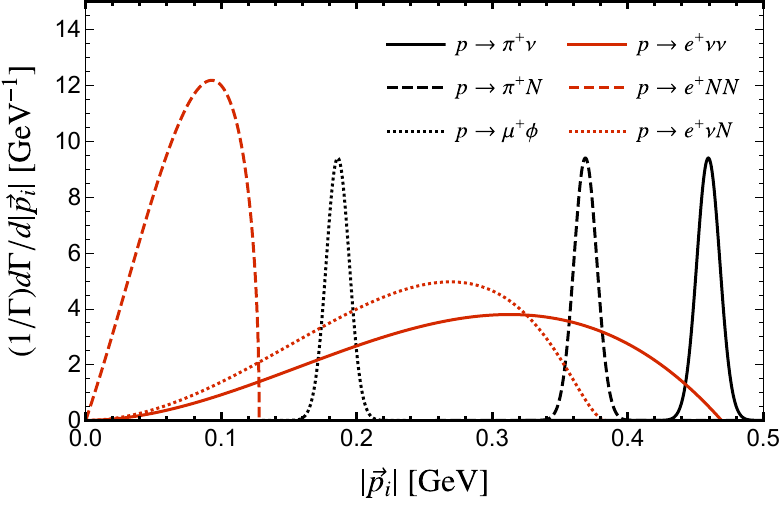}
    \caption{Normalized distribution of visible particle momentum $ |\vec p_i |$ for $i\in\{\pi^+,e^+,\mu^+\}$ in two- and three-body decays, for a sterile neutrino mass $m_N=400$~MeV and dark scalar mass $m_\phi=700$~MeV. The two-body distributions have been multiplied by a factor $\tfrac{1}{5}$ for visibility.}    \label{fig:pion_kinematics}
\end{figure}

In Fig.~\ref{fig:pion_kinematics}, 
we compute and illustrate distribution of visible particle momenta of $\pi^+$ (solid black), $e^+$ (solid red) for
2-body and 3-body decay modes, $p\rightarrow \pi^+\nu$ (searched in Ref.~\cite{Super-Kamiokande:2013rwg}) and $p\rightarrow e^+\nu\nu$ (searched in Ref.~\cite{Super-Kamiokande:2014pqx}), respectively. The importance of nucleon decay momenta distributions, in case of conventional nucleon decays with SM constituents in the final state, was highlighted in~Ref.~\cite{Chen:2014ifa,Heeck:2019kgr}.

We display in Fig.~\ref{fig:pion_kinematics} (dashed and dotted black and red lines) NCNDK modes that provide similar visible final state signatures as nucleon decays with SM final states, but with dramatically distinct peaked momenta distributions~\footnote{Note that these distributions are normalized to have the same total widths, while for more massive final state particles there is an additional phase space suppression that lowers the decay rate. However, unless the new particle is close to being degenerate with the parent nucleon, this suppression is generally less than an order of magnitude, and could be compensated for by a small change in the new-physics scale.}. 
Here, we assumed $m_N=400$~MeV and $m_\phi=700$~MeV.
We note that the effects of the Fermi motion (e.g.\ Ref.~\cite{Nakamura:1976mb}), nuclear binding energies (e.g.\ Ref.~\cite{Super-Kamiokande:2017gev}) as well as predicted nucleon-nucleon correlation in decays~\cite{Yamazaki:1999gz} can contribute to the modification of the momenta distributions for nucleon decays in nuclei.

Modified momenta associated with NCNDKs can significantly impact conventional nucleon decay searches, which can misinterpret or even completely miss such processes as we discuss.
The Super-Kamiokande spectral search for $p \rightarrow e^+\nu\nu$~\cite{Super-Kamiokande:2014pqx} as well as $p \rightarrow e^+X$ (here $X$ was assumed massless)~\cite{Super-Kamiokande:2015pys} considered a sample with imposed visible reconstructed $e^+$ momentum cut of 100~MeV$<p_e < 1000$~MeV.
For $p \rightarrow e^+\nu\nu$, $97\%$ of true visible signal momentum distribution lies above 100~MeV. On the other hand, see Fig.~\ref{fig:pion_kinematics} examples, for sterile neutrino with mass $m_N = 400$~MeV for NCNDK $p \rightarrow e^+NN$ only $26\%$ of true visible momentum distribution exceeds 100 MeV. However, in case of NCNDK $p \rightarrow e^+\nu N$, such $m_N$ gives $95\%$ of signal momentum distribution above the cut. Hence, signal efficiency can be dramatically modified in NCNDKs leading to misinterpretation of signals and with lifetime limits altered by over a factor of a few.

More crucially, as we illustrate in the case of sterile neutrino $m_N = 420$~MeV, the channel $p \rightarrow e^+NN$ is completely invisible in the search as visible $e^+$ momentum is always below 100~MeV. Similarly, for 2-body channel $p\rightarrow e^+\phi$ and dark scalar $\phi$ mass $m_{\phi} = 840$~MeV the peaked visible $e^+$ momentum is also completely below 100~MeV. While for $m_{\phi} = 700$~MeV the $e^+$ peak is visible above 100~MeV, but mostly invisible above a similar muon-specific cut of 200~MeV for $p\rightarrow \mu^+\phi$~\cite{Super-Kamiokande:2015pys}. Hence, conventional searches can completely miss such physics.

We note that even if the whole relevant visible momentum range of NCNDK is accessible, a shift to lower momentum distribution due to light new physics compared to SM final state search can lead to distinct results. This is possible, for example, if the uncertainties or signal efficiencies are not uniform over momenta distributions.  

In dinucleon or trinucleon decays, higher available maximum energies allow to probe new light physics final state constituents above GeV in masses and unlocking additional accessible parameter space.

\paragraph{\textbf{Conclusions.} --- } 
Nucleon decays have long been sought as key signatures of motivated fundamental theories. We demonstrated that non-canonical nucleon decays offer a unique window of opportunities for broad exploration of light new physics beyond the SM, including dark photons, ALPs, sterile neutrinos, and scalar dark sectors. As we showed, conventional nucleon decay searches can not only misinterpret but also completely miss such new physics. We propose a general strategy based on momentum distribution of decays, especially with invisible final states, which offers a rich program connecting distinct subfields to search for new physics through non-canonical nucleon decays in existing as well as upcoming major experiments such as Super-Kamiokande, Hyper-Kamiokande, DUNE, and JUNO. 
\begin{acknowledgments}
\paragraph{\textbf{Acknowledgments.}---}
We thank Hank Sobel for discussions. K.F. acknowledges support from the JSPS KAKENHI grants JP21H01086 and JP22K21350.
C.H. acknowledges support from the CDEIGENT grant No.
CIDEIG/2022/16 (funded by Generalitat Valenciana under Plan Gen-T), and partial support from the Spanish grants PID2020-
113775GB-I00 (AEI/10.13039/501100011033), and Prometeo CIPROM/2021/054 (Generalitat Valenciana). C.H. also acknowledges support from the IISN convention No. 4.4503.15.
V.T. acknowledges the support by the World Premier International
Research Center Initiative (WPI), MEXT, Japan, and the JSPS KAKENHI grant 23K13109.
\end{acknowledgments}

\appendix

\section{Appendix A: 2-body and 3-body Nucleon Decays} 
To calculate the rate of nucleon decay via the operators in Tab.~\ref{tab:op_list} we use the nucleon form factor~\cite{Aoki:2017puj}
\begin{equation}
\begin{aligned}
\label{eq:formfactorPion}
    &\bra{\pi}\epsilon_{abc}\left(q^aCP_\Gamma q^b\right)P_{\Gamma'}q^c\ket{\psi} =\\
    &\qquad P_{\Gamma'}\Big[W_0^{\Gamma\Gamma'}(\mu,p^2) -\frac{i\slashed p}{m_\psi}W_1^{\Gamma\Gamma'}(\mu,p^2)\Big]u_\psi\, ,
\end{aligned}
\end{equation}
where $\psi\in \{n,p\}$ is a nucleon, and $q\in \{u,d\}$ is a quark~\footnote{Note, however, that some combination of quarks may not lead to nucleon decay with a final state pion. This form factor is not applicable in such cases.}. Furthermore, $p$ is the transferred momentum, $\mu$ is the energy scale, and $\Gamma,\,\Gamma'\in\{L,R\}$ denote left- or right-handed chiralities. For convenience we also define the matrix element for a purely right-handed operator as
\begin{equation}
\begin{aligned}
\label{eq:formfactorPionRR}
    &\bra{\pi}\epsilon_{abc}\left(q^aCP_R q^b\right)P_Rq^c\ket{\psi} \equiv F^\pi_\psi(\mu,p^2)u_\psi\, .
\end{aligned}
\end{equation}
We now turn to the decay $p\to\pi^+ N$ as an example, which may be induced by $\mathcal{O}_{d^2 u N}$ from Tab.~\ref{tab:op_list}. The matrix element can be written as
\begin{equation}
    \mathcal{M}_{p\to\pi^+ N}=\bra{N\pi^+}C_{d^2 u N}(\mu_\text{NP})\mathcal{O}_{d^2 u N}\ket{p} \, ,
\end{equation}
where, using Eq.~\eqref{eq:formfactorPionRR}, we have
\begin{equation}
\begin{aligned}
    &\mathcal{M}_{p\to\pi^+ N}=\\
    &\qquad U'(\mu_\text{NP},\mu_0)C_{d^2 u N}(\mu_\text{NP})F_p^{\pi^+}(\mu_0,m_N^2)u_pP_R\bar u_N \, .
\end{aligned}
\end{equation}
As another example, we consider the decay $p\to e^+\phi$ mediated via $\mathcal{O}_{du^2e\phi}$ from Tab.~\ref{tab:op_list}. The matrix element can be written as
\begin{equation}
\begin{aligned}
    &\mathcal{M}_{p\to e^+\phi}=\bra{e\phi}C_{d e u^2 \phi}(\mu_\text{NP})\mathcal{O}_{d e u^2 \phi}\ket{p}\, ,
\end{aligned}
\end{equation}
where the form factor is given by~\cite{Aoki:2017puj}
\begin{equation}
\label{eq:formfactor_vacuum}
    \bra{0}\epsilon_{abc}\left(q^a P_\Gamma q^b\right) P_{\Gamma'} q^c\ket{\psi} = \alpha_\psi^\Gamma(\mu) P_{\Gamma'} u_\psi\, .
\end{equation}
Using Eq.~\eqref{eq:formfactor_vacuum} we have
\begin{equation}
\begin{aligned}
    &\mathcal{M}_{p\to e^+\phi}=\\
    &\qquad U'(\mu_\text{NP},\mu_0)C_{d e u^2 \phi}(\mu_\text{NP})\alpha_p^R(\mu_0) u_p P_{R} \bar{u}_e
\end{aligned}
\end{equation}

From these two-body decays, we can calculate the decay width as
\begin{equation}
\begin{aligned}
    \Gamma_{\psi\to i j} = \frac{1}{16\pi}\frac{\lambda^{1/2}(m_\psi,m_{i},m_{j})}{m_\psi^3}|\mathcal{M}_{\psi\to i j}|^2 \, .
\end{aligned}
\end{equation}
Here the momentum $\vec p_i$ of the final state particles $i$ and $j$ have opposite directions and the same magnitude $|\vec p_i|$ given by
\begin{equation}
    |\vec p_i| = \frac{\lambda^{1/2}(m_\psi^2,m_i^2,m_j^2)}{2m_{\psi}}\, .
\end{equation}

The differential rate for the three-body $\psi\to i j k$ mode with respect to the momentum one one of the final state particles is given by
\begin{equation}
    \frac{d\Gamma_{\psi\to i j k}}{d|\vec p_i|}=\frac{ds}{d|\vec p_i|}\frac{d\Gamma_{\psi\to i j k}}{ds}
\end{equation}
where $s\equiv (p_{\psi}-p_i)^2$ and where in the centre-of-mass frame we have
\begin{equation}
    \frac{ds}{d|\vec p_i|}=-\frac{2m_\psi |\vec p_i|}{\sqrt{m_i^2+|\vec p_i|^2}}\, .
\end{equation}
The differential decay rate with respect to $s$ is given by
\begin{equation}
    \frac{d\Gamma_{\psi\to i j k}}{ds}=\frac{1}{(2\pi)^3}\frac{1}{32m_\psi^3}\int_{t^-}^{t^+} dt |\mathcal{M}_{\psi\to i j k}|^2
\end{equation}
for $t\equiv (p_\psi-p_j)^2$, where
\begin{equation}
\begin{aligned}
t^\pm =&~ \frac{(m_{\psi}^2-m_{i}+m_j^2-m_k^2)^2}{4s}\\
&-\frac{1}{4s}\left(\lambda^{1/2}({s,m_\psi^2,m_i^2})\mp \lambda^{1/2}({s,m_j^2,m_k^2})\right)^2\, .
\end{aligned}
\end{equation}

\section{Appendix B: Running of Effective Interactions} 
The running of the strong coupling $\alpha_S$ between the scale of the nucleon decay $\mu_0$ and the scale at which the operator is generated $\mu_\text{NP}$ is incorporated by the function $U'(\mu_\text{NP},\mu_0)$. Assuming $\mu_\text{NP}>m_t$ and $m_b>\mu_0>m_c$ it is given by
\begin{equation}
\begin{aligned}
    &U'(\mu_\text{NP},\mu_0)=\\
    &\quad U_i^{N_f=6}(\mu_{NP},m_t) U_i^{N_f=5}(m_t,m_b) U_i^{N_f=4}(m_b,\mu_0)
\end{aligned}
\end{equation}
where~\cite{Aoki:2008ku}
\begin{equation}
\begin{aligned}
	\label{eq:runningfactor}
	&U_i^{N_f}(\mu_1,\mu_2) = \lr{\frac{\alpha_S(\mu_2)}{\alpha_S(\mu_1)}}^{\gamma_i^{0}/2\beta_0}\times \\
 &\qquad \Bigg[1+\lr{\frac{\gamma_1}{2\beta_0}-\frac{\beta_1\gamma_0}{2\beta_0^2}}\frac{\alpha_S(\mu_2)-\alpha_S(\mu_1)}{4\pi}\Bigg]\, .
\end{aligned}
\end{equation}
For the operators given in the examples above we have
\begin{align}
    &\beta_0=11-\frac{2}{3}N_f,\; \beta_1=102-\frac{38}{3}N_f,\\
    &\gamma_0=-4,\; \gamma_1=-\left(\frac{14}{3}+\frac{4}{9}N_f\right)\, ,
\end{align}
where $N_f$ is the number of fermions.

\bibliographystyle{utphys3}
\bibliography{biblio}

\end{document}